# Cryogen-free one hundred microKelvin refrigerator

Jiaojie Yan[1], Jianing Yao[1], Vladimir Shvarts[2], Rui-Rui Du[1, 3, 4], and Xi Lin[1, 3, 4, *]

[1]*International Center for Quantum Materials, Peking University, Beijing 100871, China*
[2]*Janis Research Co., Woburn, MA 01801-2025, USA*
[3]*Beijing Academy of Quantum Information Sciences, Beijing 100193, China*
[4]*CAS Center for Excellence in Topological Quantum Computation, University of Chinese Academy of Sciences, Beijing 100190, China*

\* xilin@pku.edu.cn

**Abstract:**

*Temperature below 100 µK is achieved in a customized cryogen-free dilution refrigerator with a copper-nuclear demagnetization stage. The lowest temperature of conduction electrons of the demagnetization stage is below 100 µK as measured by a pulsed platinum NMR thermometer and the temperature can remain below 100 µK for over 10 hours. An up to 9 T demagnetization magnetic field and an up to 12 T research magnetic field can be controlled independently, provided by a coaxial room-temperature-bore cryogen-free magnet.*

## I. INTRODUCTION

MilliKelvin temperature environments are crucial conditions for observing exotic quantum phenomena such as fragile fractional quantum Hall states[1-6] and novel quantum phase transitions[7, 8] which require thermal excitations smaller than their characteristic energy scales. It is reasonable to expect that more vulnerable quantum phases[9-11] and unexpected new physics should be investigated in a lower temperature regime. In addition to fundamental physics, well-understood problems technically benefit from lower temperatures as well, such as longer coherence time for semiconductor qubits and higher fidelity for superconducting qubit operations. Although milliKelvin refrigerators have been commercially available and popular in research universities and institutes, access to microKelvin temperature regime is still limited worldwide and beyond commercial solutions.

To reach milliKelvin regimes, techniques involving evaporating liquid $^3$He, adiabatic demagnetization of paramagnetic salts, and $^3$He – $^4$He dilution refrigeration are used to achieve typical temperatures of 300 mK, 50 mK, and 10 mK. Towards extremely low temperature conditions for condensed matter, methods such as Pomeranchuk cooling[12, 13] and adiabatic nuclear demagnetization[14] have been developed. The latter method, which is used in this work, is widely applied in the microKelvin regime and has reached the lowest environmental temperature so far. The principle of adiabatic demagnetization depends on the fact that the occupation probabilities of spin energy levels are determined by the ratio of magnetic field and temperature, so that temperature decreases when the magnetic field is lowered in an adiabatic process. Nuclear spins have been cooled to 100 pK[15] while environmental temperature for refrigerators can be cooled to about 10 µK.[16] There are dozens of adiabatic nuclear demagnetization refrigerators worldwide, located in the USA,[17, 18] Europe,[19] Japan,[20, 21] and other places,[22] allowing researchers to investigate interesting quantum systems, including two-dimensional electron gas,[23, 24] superconductivity,[25, 26] and helium physics.[27-30]

Conventional cryostats below 4 K were mostly pre-cooled by liquid $^4$He. However, a revolutionary era of replacing conventional cryostats with their cryogen-free counterparts is progressing. With the worldwide shortage of liquid $^4$He and rocketing price for liquid $^4$He,[31] applications of cryogen-free techniques[32] have turned cryogenics into an ordinary tool friendly to users without the need of transferring liquid $^4$He. Therefore, cryogen-free cryostats are also called dry cryostats. The cryogen-free systems based on Gifford-McMahon refrigerators or pulse tube refrigerators (PTRs) are able to provide sustainable low temperature environments with large accessible spaces, albeit with additional technical challenges such as extra vibrations, lower cooling power at 4 K, and worse temperature stability compared with refrigerators pre-cooled by liquid $^4$He. Furthermore, spared of liquid $^4$He transfer, cryogen-free systems reveal the potential to perform long-period automatic operations, and are becoming better candidates of low temperature environments for complex ambient conditions, quantum computers, and outer space applications. Adiabatic nuclear demagnetization based on cryogen-free cryostat has attracted interest from a number of research groups as well as us over the past decade, and at least three systems have already been reported.[33-35]

In this paper, we present a cryogen-free adiabatic nuclear demagnetization system that can cool below 100 μK. The demagnetization stage, consisting of high purity copper, is precooled by a dilution refrigerator at a constant magnetic field over 8 T. After the magnetic field was reduced to 10 mT, the refrigerator temperature was measured to be less than 100 μK with a pulsed platinum NMR thermometer. The demagnetization stage can stay below 100 μK for over 10 hours. The detailed design for the components and procedures of preparation are provided in this work, which could be a convenient starting point for constructing standard cryogen-free microKelvin experimental platforms.

**II. THE CRYOGEN-FREE DEMAGNETIZATION SYSTEM**

The cryogen-free demagnetization system consists of a customized dilution refrigerator and a room-temperature-bore superconducting magnet. The refrigerator and the magnet reside in two separate vacuum chambers and are cooled by two separate PTRs. Taking advantage of low nuclear ordering temperature,[36] excellent thermal conductance, convenient accessing and easy machining, copper is chosen as our demagnetization refrigerant. Considering copper's small electrical resistivity and corresponding eddy current in a magnetic field, accommodation and vibration damping solutions have been carefully designed for this cryogen-free system (Fig. 1). The refrigerator and the magnet are bolted to a customized pneumatic damping table, which is sited on a 0.5 m thick concrete stage extending from the ground, allowing the refrigerator to stay above the pit. The frame of the concrete stage is built of nonmagnetic 304 stainless steel. In addition, the $^3$He – $^4$He mixture pumping line passes through a sandbox hanging from the roof for vibration damping and the pumping line is equipped with bellows along two dimensions to decouple the refrigerator from circulation pumps. Other gas pipelines are also connected to the refrigerator flexibly to release tension. Inside the refrigerator, highly-flexible copper braids are used to avoid rigid mechanical connections between cold plates of the refrigerator and stages of cold head, while providing a good thermal connection. Furthermore, a Joule-Thomson unit has been installed to reduce mixture return pressure.[37] Without floating the pneumatic damping table, the vibration on the top flange of the refrigerator was once measured with an accelerator, indicating the root mean square of velocity and displacement were about 1 μm/s and 1 μm in the vertical direction.

Major components inside the refrigerator are illustrated in Fig. 2(a) with thermometers presented. The refrigerator is based on a customized dilution refrigerator provided by Janis Research Company with a 1.58 m tail (JDry-500 NADRA). The refrigerator has five cold plates with different temperatures, named as 50 K (~ 47 K), 3 K (~ 3.2 K), still (~ 800 mK), intermediate (~ 60 mK), and mixing chamber (MC) plates (~ 10mK). Radiation shields are thermally anchored at 50 K, 3 K, and intermediate plates. The 3 K all-copper indium-sealed shield (named as IVC shield in this work) is made vacuum tight and light tight, so that the dilution unit and demagnetization components can be pre-cooled by exchange gas. This refrigerator offers 24 coaxial cables and 32 pairs of twisted superconducting wires from room temperature to the MC plate for measurement. Besides, 6 pairs of twisted copper wires from room temperature to the 3 K plate provide opportunities to pass through 3 different loops of large current, such as currents for superconducting solenoids of NMR thermometers and heat switch (HS) via superconducting wire extensions.

Below the dilution unit the copper thermal link (TL) is located, aiming to reduce the impact of the demagnetization solenoid stray field on the MC plate. The HS is mounted below the TL, capable of tuning the thermal conductance between the TL and the demagnetization stage. The HS has nine bent aluminum foils (0.2 mm thick and ~ 10 mm wide each) bolted together with silver sheets, which are welded to silver flanges mounted on either the TL or the demagnetization stage (Fig. 2(b) and Fig. 2(c)). The contact surfaces of aluminum foils are gold plated with a zinc underlayer. A special shaped superconducting magnet with solenoid and rectangular part in series surrounds the aluminum foils, able to provide magnetic fields $B > B_c = 10.5$ mT at a current of 0.4 A that can break down the superconducting state in aluminum. The solenoid is surrounded with a niobium shield for better magnetic field homogeneity and to shield against the disturbance of the demagnetization field. In the normal state, heat can be easily transmitted from the demagnetization stage to the TL by electrons, which is defined as the "on" state for HS. However, in the superconducting state, electrons form Cooper pairs and the thermal conductivity decays exponentially with temperature, which is defined as the "off" state for HS.[38] For better thermal isolation in the "off" state, the supporting structure between the TL and the demagnetization stage consists of stainless steel tubes in company with 10 mm thick Vespel SP-22 gaskets. The demagnetization stage is 743 mm in height, with a maximum diameter of 68 mm (Fig. 2(d)), weighing 11.7 kg (184 mol) of copper. Copper for the demagnetization stage is of nominal 99.999% purity (Cu + Ag > 99.999% in fact) and annealed at 950 °C for 60 h in a ~$10^{-3}$ mbar vacuum inside a homemade quartz tube furnace. Crystalline grains can be seen on the surface (Fig. 2(e)) after annealing, indicating the high purity and appropriate stress relaxation of copper. Both the TL and the demagnetization stage are made of single copper cylinders machined with 1.0 mm grooves, in order to reduce eddy currents during magnet field ramping.

The lower part of the refrigerator, located in the center of the 12 T research magnetic field, is designed for the sample holder. Refrigerator shields are designed consisting of three components: a wide and short can with a hole on the bottom, a narrow and long can with a hole on the bottom for the demagnetization stage, and a short tail inserting into the center of the 12 T magnetic field. Changing samples will not require disassembling the full set of shields, but only the bottom part of each shield.

The superconducting magnet from Cryomagnetics, Inc.[39] which provides a room-temperature bore with a separate vacuum and PTR from the refrigerator, is sited in a nonmagnetic customized elevator. The refrigerator is secured when moving the elevator and disassembling shields. When the magnet is mounted together with the refrigerator, the elevator can stay or be moved away. Controlled by independent power supplies, two separate solenoids provide individual magnet field for the demagnetization stage and the position of the sample holder. Magnetic field distribution with both solenoids in full currents is shown in Fig. 2(f). The demagnetization field is within 9 T and the region of $B > 8$ T is 25 cm in height, corresponding to ~ 95 mole copper in the demagnetization stage. The maximum research field is 12 T, larger than other reported cryogen-free adiabatic demagnetization refrigerators. Two compensation regions ($B < 10^{-2}$ T) aiming to reduce the boundary thermal resistance in the magnetic field, and the magnetic field's impact on the HS and thermometers, are located at both ends of the demagnetization stage.

With all components for demagnetization attached, the refrigerator can cool down to its dilution base temperature ~ 8.6 mK from room temperature in about 45 hours (including 6 hours for exchange gas pumping). No liquid nitrogen is transferred into the refrigerator but a mechanical moveable heat switch connecting the 50 K plate and the 3 K plate accelerates the initial cooling from room temperature. Exchange gas is also introduced into IVC at room temperature to speed up initial cooling for components below the 3 K plate. The cooling power from 10 mK to 140 mK is shown in Fig. 2(g). With extra components for demagnetization attached, the refrigerator performs similarly with the bare dilution refrigerator, which is expected to be 13 μW at 20 mK and 500 μW at 100 mK.

**III. THERMOMETRY**

Thermometers of different working ranges are mounted to key positions of the refrigerator. Calibrated Cernox thermometers are mounted on the 50 K plate, 3 K plate and the bottom of IVC shield,[40] while $RuO_2$ thermometers are mounted on the still, intermediate, MC plates, TL and the demagnetization stage.[41] $RuO_2$ thermometers are susceptible to electromagnetic interference and self-heating, especially at ultra-low temperatures, so an extra cerium magnesium nitrate (CMN) sensor and a superconducting fixed-point device (FPD) are mounted adjacently on the TL, for checking the stability of resistor thermometers. Temperature measurement in milliKelvin and microKelvin regimes, which is the temperature range of nuclear adiabatic refrigeration, is challenging. Excellent thermal contact and very low heat dissipation should be satisfied for thermometers at these temperatures. Elaborate thermometers taking advantage of vibration wire resonance in superfluid $^3$He,[42] thermal noise,[43] and nuclear magnetic susceptibility[38] have been developed to work in these temperature regimes. In this work, an in situ $^3$He melting pressure thermometer (MPT)[44] and a pulsed platinum NMR thermometer[45] are applied to establish our thermometry.

Presently there exists no formal international temperature scale below 0.65 K. A provisional temperature scale based on $^3$He melting pressure named PLTS-2000[46] is used as our temperature scale above 0.902 mK. Our MPT is a Straty-Adams type pressure gauge[47] illustrated in Fig. 3(a). Silver powders are squeezed inside the MPT cell in order to enhance the thermal conductance between $^3$He and the silver base, which promotes the response of the MPT.[48] When the pressure of the $^3$He in the cell changes, the diaphragm deforms elastically, reflecting on the capacitance $C_p$. Pressure can be determined from the ratio $C_p/(C_p+C_{ref})$, measured by a seven-decade ratio standard[49] together with a lock-in amplifier in a bridge circuit (Fig. 3(b)).[44] The CuNi capillaries are thermally anchored at 3 K, still, intermediate, MC plates and TL before being connected to the cell.

The MPT cell is filled with liquid $^3$He and calibrated at 1.2 K with a room temperature quartz pressure transducer.[50] With temperature decreasing at a constant pressure of 3.47 MPa, $^3$He in the cell becomes a solid-liquid coexistence phase and follows the melting curve. Comparisons between the MPT and a calibrated $RuO_2$ thermometer from 10 mK to 560 mK are shown in Fig. 3(c). For $T > 30$ mK, $T_{RuO_2}$ and $T_{MPT}$ are consistent, while for $T < 30$ mK, the resistor thermometer begins to saturate and $T_{RuO_2}$ deviates from $T_{MPT}$. The Néel transition, superfluid A-B transition, and superfluid A transition for $^3$He can be observed from MPT, which are shown in Fig. 3(c) (right panels from top to bottom). The Néel transition, the ordering transition in solid $^3$He, can be recognized as an obvious slope alteration after the transition and a less obvious pressure plateau. A pressure plateau can be observed during the superfluid A-B transition as well, and the superfluid A transition is a second-order transition with a slope kink in $P(t)$. Reference capacitor in the MPT shown in Fig. 3(a) is more stable than a room temperature reference capacitor, although the latter can still be used in a measurement of Néel phase transition.

The lowest temperature of PLTS-2000 is limited by $T_{Néel} = 0.902$ mK, lower than which the melting pressure becomes much less sensitive to temperature changes. Therefore the platinum NMR thermometer is applied for $T < 0.9$ mK. $^{195}$Pt nuclei possess magnetic moment and the magnetic susceptibility $\chi$ obeys paramagnetic temperature dependence in microKelvin regime in approximation. The deviation from $1/T$ for $^{195}$Pt at $B = 26.8$ mT (the case in this work) is 0.1% at 100 μK in principle. Moreover, taking advantage of pure platinum sensor, the NMR thermometer can be in excellent thermal contact with the demagnetization stage because of electron conduction through the metal-metal interface and strong spin-lattice coupling within platinum.[38]

Fig. 4(a) presents a photograph of our NMR thermometer (with the solenoid disassembled). In our measurement, single-coil setup with PLM-5[51] is applied. Few hundred 25 micron bare platinum wires form a brush, which is welded to a silver rod bolted to the top flange of the demagnetization stage. The illustration of thermometer structure is shown in Fig. 4(b). The solenoid generates a static magnetic field $B_z$ which polarizes the nuclear spins of platinum along z-axis at $I = 0.5371$ A. Then resonance pulses at 235.4 kHz are applied by the coil along the x-axis, tipping nuclear magnetization away from its original vertical orientation. The projection of nuclear magnetization can be picked up from the same coil after each excitation pulse. A schematic of an applied excitation pulse and a picked up free induction decay (FID) signal is shown in the inset of Fig. 4(c). Excitation pulses applied are 2.969 $V_{pp}$ in amplitude with durations varying from 63.7 μs (15 cycles) to 17.0 μs (4 cycles). Excitation power decreases as temperature decreases, accelerating the equilibration time through lowering tipping angles of magnetization, and reducing self-heating. Measurement circuit is shown in Fig. 4(c), where the capacitance in the shaded region represents the parasitic capacitance.

NMR thermometers are secondary thermometers which require at least one fixed point for calibration according to $M = C_0/T$. Here $M$ is a value of arbitrary unit extracted from the FID signal that is proportional to magnetic susceptibility χ, and $C_0$ is a constant to be determined. In principle, all three $^3$He phase transitions can be employed as temperature fixed points and the entire $^3$He melting curve can be used to cross-check the NMR thermometer. However, superfluid A transition is hard to be identified while superfluid A-B transition could be shifted by the supercooling effect. And the NMR thermometer has better signal-to-ratio at lower temperatures. As a result, the NMR thermometer was calibrated using Néel transition, the lowest temperature phase transition and the most distinguishable one, in a cooling down process. When the temperature approached $T_{Néel}$, the ramping rate was slowed to 0.04 mT/s. The field sweeping was paused when slope alteration occurred in $P(t)$. After the NMR thermometer reached thermal equilibrium with the demagnetization stage, we defined $T_{NMR} = T_{Néel} = 0.902$ mK. Inevitably, the actual temperature of the demagnetization stage in this process should be a little bit lower than $T_{Néel}$, indicating the actual temperature of the demagnetization stage should be lower than what we report.

**IV. DEMAGNETIZATION STAGE PERFORMANCE**

Adiabatic nuclear demagnetization provides a single-shot method for refrigeration to microKelvin regime, which includes two major steps. The first step is to cool the demagnetization stage to dilution base temperature with the HS at "on" condition and apply external magnetic field $B_{start}$. The heat of magnetization is gradually (usually 4 to 7 days for us) removed by the dilution refrigerator until demagnetization stage temperature $T_{start}$ is typically below 20 mK. The second step is to slowly decrease the applied magnetic field to $B_{final}$, while switching the HS to "off" state when the temperature of the demagnetization stage approaches to that of the MC plate. If this process is done slowly enough, it can be considered as an adiabatic process. Therefore, the constant entropy $S$ of copper nuclear spins, which means the constant ratio of $B/T$, leads to a temperature decrease. This in turn cools conduction electrons in copper, and temperature sensors. Fig. 5(a) presents the results of such a process with $B_{start} = 8.5$ T and $T_{start} = 17.7$ mK as initial conditions reached after 4.2 days of pre-cooling. The field was ramped down at the following rates of 0.1 mT/s from 8.5 T to 5.3 T, 0.2 mT/s to 2.1 T, 0.08 mT/s to 0.47 T and 0.04 mT/s to 10 mT. The final field of 10 mT was kept constant for the rest of the cool down with the power supply permanently applying $I = 0.1$ A, because there is no persistent mode in the magnet. The HS was switched to "off" at $T_0 = 11.1$ mK and $B_0 = 5.3$ T and the value $B_0/T_0$ is used to predict the temperature according to the ideal adiabatic process and calculate the efficiency $\xi = (B/T)/(B_0/T_0)$. As shown in the inset of Fig. 5(a), the efficiency drops below 80% sharply at $B < 0.2$ T. This may be attributed to the eddy current heating in copper as well as residual heat releases to the demagnetization stage.

Temperature of the demagnetization stage is shown in Fig 5(b) as a function of time. After reaching the final field $B_{\text{final}} = 10$ mT, it took about 15.3 hours to achieve the lowest temperature $T_{\text{final}} = 95$ μK. Such a long time could be caused by the fact that thermal equilibrium between $^{195}$Pt nuclei and thermal reservoir of copper nuclei in demagnetization stage is mediated by conduction electrons in both materials. The spin-lattice relaxation time $\tau_1$ follows Korringa law $\tau_1 T_e = \kappa$ [52] and can reach an hour in copper at 0.3 mK.[38] Another demagnetization started at $B_{\text{start}} = 8.9$ T and $T_{\text{start}} = 18.6$ mK reached 90 μK as shown in Fig. 5(c). The demagnetization stage remained below 100 μK for more than 10 hours in both demagnetizations.

The increase in temperature after one demagnetization as a function of time is presented in Fig. 6. The heating power $\dot{Q}$ warming up the demagnetization stage can be determined by the linear fitting of $1/T$ versus time. Here the molar heat capacity of the demagnetization stage is dominated by copper nuclear spins $C_n = \frac{\lambda_n B^2}{\mu_0 T_n^2}$ ($B \gg B_{\text{internal}} = 0.36$ mT) in a magnetic field $B$, where $\lambda_n$ is the molar nuclear Curie constant, $\mu_0$ is the vacuum permeability ($\lambda_n/\mu_0 = 3.22$ μJ·K·T$^{-2}$·mol$^{-1}$ for copper[38]) and $T_n$ is the temperature of nuclei. Supposing $\dot{Q}$ is fully used to warm up the nuclear spins via hyperfine interactions, one gets $\frac{d}{dt}\left(\frac{1}{T}\right) = -\left(\frac{n\lambda_n B^2}{\mu_0 \dot{Q}} + \kappa\right)^{-1}$.[35] Here $T$ is the measured electron temperature, and $\kappa$ is the Korringa constant independent of temperature taken as 1.2 K·s.[38, 53] At $B = 10$ mT, the corresponding heat for each mole copper $\dot{Q}/n$ is derived to be 22 pW/mol from 90 μK to 115 μK, and it took over 80 hours for the stage to warm up to 1 mK as shown in Fig. 6. Taking account of the time before the temperature began to increase, the holding time is ~ 90 hours at 10 mT. Heat releases at different magnetic fields are shown in the inset of Fig. 6, and other values are obtained around 0.65 mK, 0.85 mK, and 1.75 mK. The biggest concern of nuclear demagnetization cooling with cryogen-free techniques is the extra heating from the eddy current caused by vibrations in the magnetic field. The structure and design presented in this work demonstrate how to minimize the heating from vibration to a level low enough for a 100 μK environment.

**V. DISCUSSIONS**

According to the relationship between $1/T$ and $t$ mentioned above, if the heating power can be considered constant, the lower initial temperature can be deduced from the time dependence of higher temperature, $T(t)$. This method provides the opportunity to measure a much lower temperature value beyond thermometers' original working range.[35, 38, 54] Such an empirical method is extremely useful in the application at ultra-low temperature, for both extending the range and checking the reliability of thermometers. However, how accurately one can measure the temperature through such an estimate was unknown.

A heating power $P_{\text{app}} = 250$ nW was applied at $B = 0.155$ T starting from $T = 1.94$ mK for the purpose of examining the accuracy of the extrapolation. In Fig. 7(a), $1/T$ (black solid line) shows a linear behavior with time at the beginning, and tends to saturate at high temperature end. The red, yellow, green, and blue solid lines represent linear fittings within different temperature ranges and their extension lines, leading to extrapolation results $T_{\text{extrapolation}}$ of 1.94 mK, 2.32 mK, 2.79 mK and 3.48 mK, respectively. The red solid line fitting of 2.00 – 4.00 mK leads to the same temperature as $T(0)$. And extrapolation from the yellow solid line (fitting of 8.00 – 10.00 mK) provides 0.38 mK deviation. For temperatures higher than the MC plate, temperature deviation of 0.85 mK and 1.49 mK are obtained from 13.00 – 15.00 mK and 18.00 – 20.00 mK. With temperature rising, heat capacity of electrons $C_e$ increases with $T$ while nuclei heat capacity $C_n$ decreases with $T^{-2}$. Moreover, heat transmitted via HS and other supporting structures would gradually violate the adiabatic assumption with increasing temperature. Therefore, deviations gradually emerge with higher temperature.

Analysis of linear fitting extrapolations for continuous temperature ranges is shown in Fig. 7(b), where the horizontal axis represents beginning temperatures while the vertical axis represents ending temperatures. Deviation temperature $\Delta T = T_{\text{extrapolation}} - T(0)$ is presented in different colors. Less than 20% deviation can be derived from fitting temperature below 10 mK, while the deviation increases rapidly for beginning temperature higher than 10 mK. This study of temperature extrapolation was started when the demagnetization stage showed obvious warming, rather than demagnetization just completed. Immediate heating after demagnetization with lower $T$ and thus larger $C_n$ might lead to better accuracy.

## VI. CONCLUSIONS

We present a cryogen-free refrigerator with a nuclear demagnetization stage that can reach temperature below 100 μK for over 10 hours. A superconducting magnet cooled by an independent PTR is fixed together with the refrigerator at room temperature, providing a 9 T maximum demagnetization field for the demagnetization stage as well as a 12 T maximum magnetic field for samples. Heat release to the demagnetization stage is about 22 pW/mol at $B = 10$ mT, and allows holding time of over 90 hours below 1 mK. The successful operation of this refrigerator reveals the potential to further develop long-operation-time microKelvin refrigerators that are completely cryogen-free, capable of automatic operation and remote control. In addition, our analysis of constant heating shows that thermometers working down to 10 mK can be used to estimate the temperature down to 2 mK within 20% deviation.

## ACKNOWLEDGEMENTS

We thank Dr. Jian-Sheng Xia for his invaluable discussions and help on the MPT as well as the demagnetization stage annealing, Prof. Xiunian Jing and Prof. Li Lu for discussions and help with NMR measurements, Dr. A.C. Clark and Dr. Zuyu Zhao for discussions, Prof. Kono Kimitoshi for discussions and the sharing. We thank Xiaohong Li, Ashley Huff and Jianmin Zheng for assistance in the lab. This work was supported by NSFC Special Fund for Research on National Major Research Instruments (Grant No. 11127902).

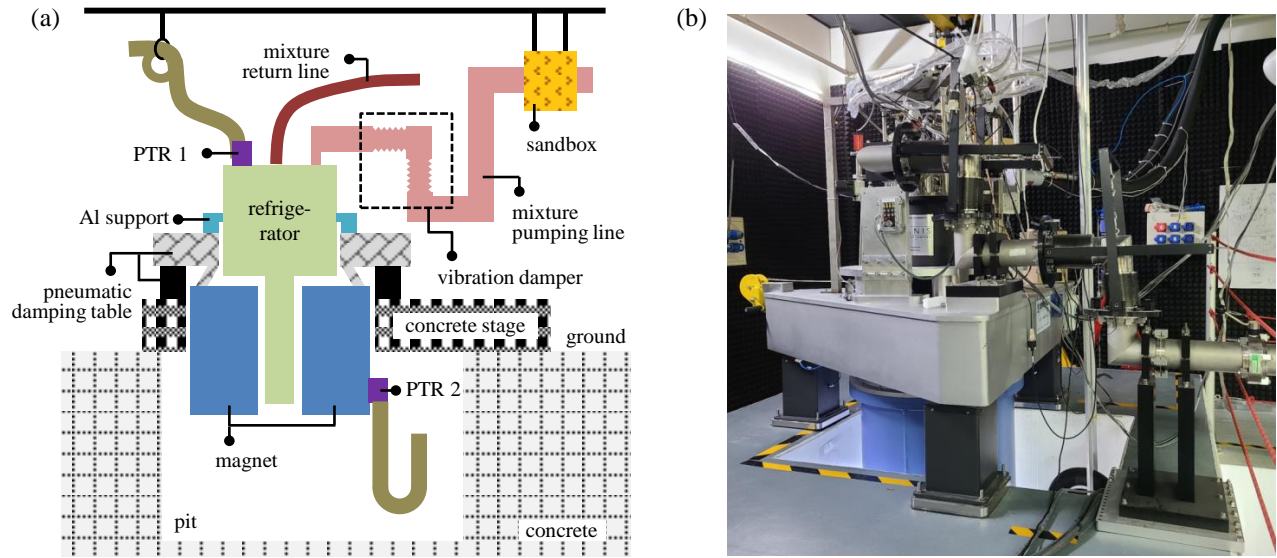

**FIG. 1.** (a) Sketch of the cryogen-free demagnetization system and vibration damping solutions. The dilution refrigerator and the room-temperature bore magnet are cooled by two independent PTRs and are mounted on a customized pneumatic damping table. The whole system is sited on a concrete stage extended from the ground to the top of the pit. (b) Photograph of the demagnetization system.

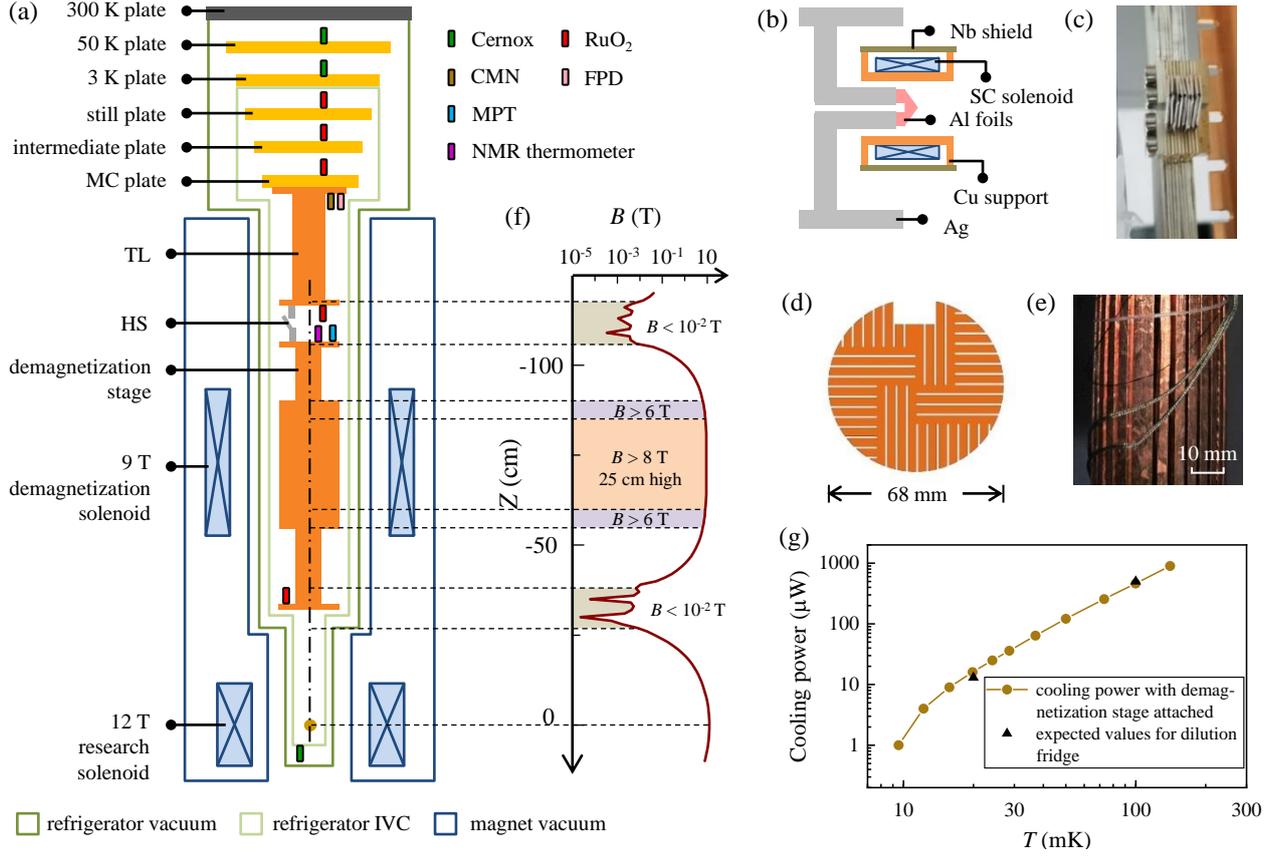

**FIG. 2.** (a) Illustration of the main components inside the refrigerator and the magnet, with thermometers in the refrigerator presented. Refrigerator components below the MC plate are drawn proportionally along the *z*-direction to match with Fig. 2(f). (b) Illustration of the HS. The copper support is thermally anchored at TL. (c) Photograph of the central part of the HS with solenoid removed. (d) The cross-sectional view of the widest region of the demagnetization stage. The grooves are 1.0 mm in width. (e) Photograph of the demagnetization stage after annealing. Copper crystalline grains can be seen on the surface. (f) Magnetic field distribution of central axis along *z*-direction with both magnet solenoids in full currents. Compensation regions ($B < 10^{-2}$ T) and high field regions ($B > 6$ T and $B > 8$ T) are shown in different colors. The expected homogeneity of 12 T field is ±0.1 % over a 10 mm diameter sphere, indicated by the dark yellow circle in Fig. 2(a). (g) Cooling power from 10 mK to 140 mK with demagnetization components attached. The expected cooling power for the Janis dilution refrigerator is 13 µW at 20 mK and 500 µW at 100 mK, labeled as black triangles.

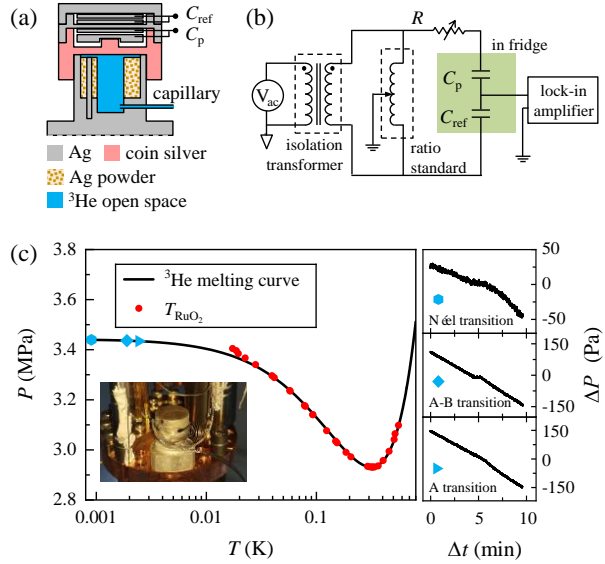

**FIG. 3.** (a) The cross-sectional sketch of the MPT. (b) Measurement circuit of the MPT. All leads are coaxial cables. Components in the refrigerator are highlighted with the green shaded region. (c) In the left main figure, black solid line is the $^3$He melting curve calculated from PLTS-2000. Red dots show the temperature of a calibrated $RuO_2$ thermometer versus MPT pressure, and light blue symbols show temperatures of the phase transitions with measured pressures. $^3$He phase transitions at $dB/dt = 0.04$ mT/s are shown on right panels. Inset shows the photograph of the MPT, which is mounted on the top plate of the demagnetization stage.

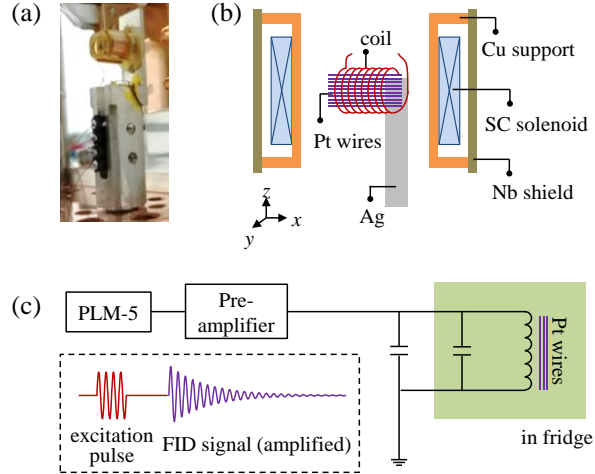

**FIG. 4.** (a) Photograph of the NMR thermometer mounted on the top plate of the demagnetization stage. Only single coil is used in measurement. (b) Illustration of the NMR thermometer. The superconducting solenoid generates a static magnetic field $B_z$, and the coil applies excitation pulses as well as picks up FID signals. The copper support is thermally anchored at the TL. (c) Measurement circuit of the NMR thermometer. Components in the refrigerator are highlighted with the green shaded region. The capacitance in the refrigerator represents the parasitic capacitance. Inset: schematic of an excitation pulse and a FID signal. In this work, the resonant frequency is 235.4 kHz, and excitation pulses applied are 2.969 $V_{pp}$ in amplitude varying from 63.7 μs (15 cycles) to 17.0 μs (4 cycles) with decreasing temperature.

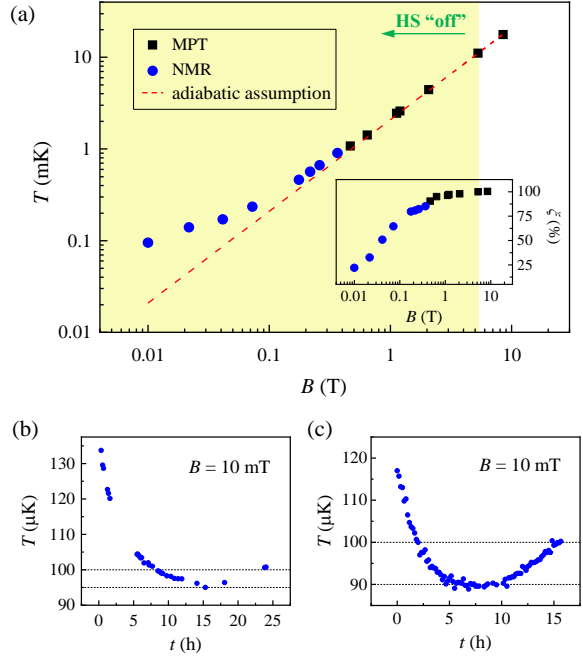

**FIG. 5.** (a) Temperature of the demagnetization stage versus the demagnetization field. $T_{\mathrm{MPT}}$ (black squares) and $T_{\mathrm{NMR}}$ (blue circles) are measured when the demagnetization is paused. The red dashed line denotes predicted temperature according to adiabatic assumption. Inset shows efficiency calculated from $\xi = (B/T)/(B_0/T_0)$. The point when the HS was switched to "off" ($T_0 = 11.1$ mK and $B_0 = 5.3$ T) is used as the reference point (100% efficiency). The adiabatic regime is highlighted with yellow shaded region in both the main panel and the inset. (b) Measured temperature as a function of time after the magnetic field setting to 10 mT, the smallest value in the described demagnetization. (c) Temperature versus time after the magnetic field setting to 10 mT in another demagnetization which reached 90 μK.

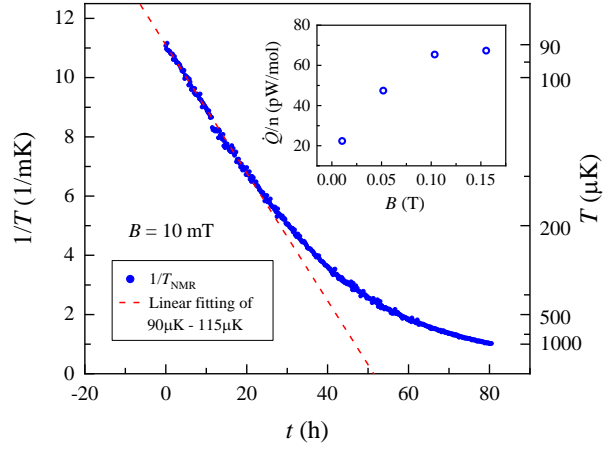

**FIG. 6.** Inverse temperature as a function of time below 1 mK at $B$ = 10 mT. It took more than 80 hours to warm up the demagnetization stage to 1 mK from 90 μK. The slope of the linear fitting (red dashed line) indicates a heat release of about 22 pW/mol. Inset: Heat release per mole at different magnetic fields. Values are calculated near 0.1 mK, 0.65 mK, 0.85 mK, and 1.75 mK.

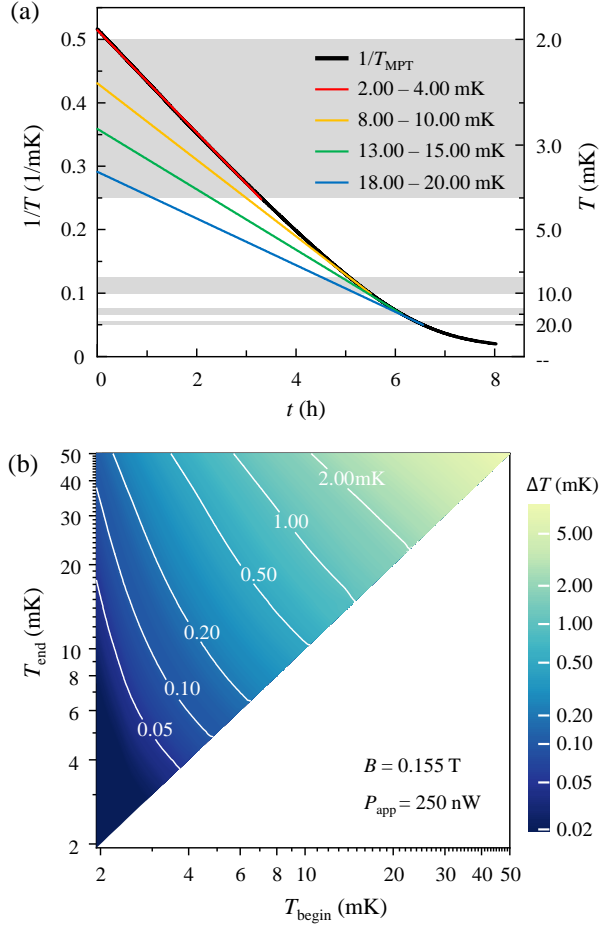

**FIG. 7.** (a) Inverse temperature as a function of time starting from $T(0) = 1.94$ mK at $B = 0.155$ T with applied heating $P_{app} = 250$ nW (black solid line). Linear fitting and extension lines of 2.00 – 4.00 mK (red), 8.00 – 10.00 mK (yellow), 13.00 – 15.00 mK (green), and 18.00 – 20.00 mK (blue) are shown and $T_{extrapolation}$ are 1.94 mK, 2.32 mK, 2.79 mK and 3.48 mK respectively. (b) Image plot of the deviation temperature $\Delta T = T_{extrapolation} - T(0)$ as a function of different beginning and ending fitting temperatures.